\documentclass[fleqn,usenatbib]{mnras}
\usepackage[T1]{fontenc}
\usepackage{ae,aecompl}
\usepackage{setspace}
\usepackage{natbib}
\usepackage{textcomp}
\usepackage{amsmath,amssymb,amsfonts,fullpage}
\usepackage{txfonts}
\usepackage{subfigure}

\usepackage[dvips]{graphicx}
\usepackage[title]{appendix}

\title{\bf{Collisional formation of massive exomoons of super-terrestrial exoplanets}}
	
\author
{		
Uri Malamud,$^{1,2}$
Hagai B. Perets,$^{2,3}$
Christoph Sch{\"a}fer,$^{4}$
Christoph Burger$^{4,5}$
\\	
$^{1}$School of the Environment and Earth Sciences, Tel Aviv University, Ramat Aviv, 6997801 Tel Aviv, Israel\\
$^{2}$Department of Physics, Technion Israeli Intitute of Technology, Technion City, 3200003 Haifa, Israel\\
$^{3}$TAPIR, California Institute of Technology, Pasadena, CA 91125, USA\\
$^{4}$Institut f{\"u}r Astronomie und Astrophysik, Eberhard Karls Universit{\"a}t T{\"u}bingen, Auf der Morgenstelle 10, 72076 T{\"u}bingen, Germany\\
$^{5}$Department of Astrophysics, University of Vienna, T{\"u}rkenschanzstra{\ss}e 17, 1180 Vienna, Austria\\
}

\date{Accepted XXX. Received YYY; in original form ZZZ}
	
\pubyear{2019}
	
\begin{document}
		
\label{firstpage}
\pagerange{\pageref{firstpage}\textendash{}\pageref{lastpage}}
\maketitle
	
\begin{abstract}
Exomoons orbiting terrestrial or super-terrestrial exoplanets have not yet been discovered; their possible existence and properties are therefore still an unresolved question. Here we explore the collisional formation of exomoons through giant planetary impacts. We make use of smooth particle hydrodynamical (SPH) collision simulations and survey a large phase-space of terrestrial/super-terrestrial planetary collisions. We characterize the properties of such collisions, finding one rare case in which an exomoon forms through a graze \& capture scenario, in addition to a few graze \& merge or hit \& run scenarios. Typically however, our collisions form massive circumplanetary discs, for which we use follow-up N-body simulations in order to derive lower-limit mass estimates for the ensuing exomoons. We investigate the mass, long-term tidal-stability, composition and origin of material in both the discs and the exomoons. Our giant-impact models often generate relatively iron-rich moons, that form beyond the synchronous radius of the planet, and would thus tidally evolve outward with stable orbits, rather than be destroyed. Our results suggest that it is extremely difficult to collisionally form currently-detectable exomoons orbiting super-terrestrial planets, through single giant impacts. It might be possible to form massive, detectable exomoons through several mergers of smaller exomoons, formed by multiple impacts, however more studies are required in order to reach a conclusion. Given the current observational initiatives, the search should focus primarily on more massive planet categories. However, about a quarter of the exomoons predicted by our models are approximately Mercury-mass or more, and are much more likely to be detectable given a factor 2 improvement in the detection capability of future instruments, providing further motivation for their development.
\end{abstract}
	

\section{Introduction}\label{S:Intro}
For the past two decades, thousands of exoplanets have been identified, providing the first detailed statistical characterization of their properties \citep{DressingCharbonneau-2013,MortonSwift-2014,BurkeEtAl-2015,MuldersEtAl-2015,FultonEtAl-2017,NarangEtAl-2018,PascucciEtAl-2018,PetiguraEtAl-2018}. However, to date, there has not been even a single confirmed detection of an exomoon.

The formation of exomoons has been relatively little studied, although they could play an important role in planet formation. Moreover, exomoon environments are important due to their potential of hosting liquid water, thereby creating more opportunities for harbouring life \citep{WilliamsEtAl-1997}, and extending the normal boundaries of what is considered habitable environments. Such a possibility is intricately contingent upon multiple factors, including the amount of insolation, tidal heating, and other heat sources available to exomoons \citep{HellerBarnes-2013,DobosEtAl-2017}, as well as their orbital stability \citep{GongEtAl-2013,HongEtAl-2015,SpaldingEtAl-2016,AlvaradoMontesEtAl-2017,ZollingerEtAl-2017,GrishinEtAl-2018,HamersEtAl-2018,HongEtAl-2018,SucerquiaEtAl-2020}, atmosphere \citep{LammerEtAl-2014,HellerBarnes-2015} and the magnetic field of either satellite or planet \citep{HellerZuluaga-2013}. Massive exomoons are also important since they can prevent large chaotic variations to their host planet's obliquity \citep{SasakiBarnes-2014}, thereby creating a more stable climate which may be essential to the survival of life on a (solid and in the habitable zone) planet \citep{NowajewskiEtAl-2018}. It has also been shown that water-bearing exomoons are in principal capable of retaining their water as their host stars go through their high-luminosity stellar evolution phases, and so they can host life or provide the necessary water for supporting life even around evolved compact stars \citep{MalamudPerets-2017}. 	

Observationally, several different ways have been proposed for the search of exomoons \citep{KippingEtAl-2009,SimonEtAl-2009,LiebigWambsganss-2010,PetersTurner-2013,Heller-2014,AgolEtAl-2015,NoyolaEtAl-2016,SenguptaMarley-2016,Forgan-2017,Lukic-2017,BerzosaMolinaEtAl-2018,VanderburgEtAl-2018}. Transit-based techniques are the most promising methods given current observational capabilities, e.g. the Hunt for Exomoons with Kepler (HEK) \citep{KippingEtAl-2012} initiative. The detectability of an exomoon chiefly relies on its orbit, mass and the mass of its host planet \citep{SartorettiSchneider-1999}. With the HEK study, exomoons are not likely to be detected below a lower mass limit of about 0.2 Earth masses ($M_\oplus$). At this mass, any exomoon would be at least one order of magnitude more massive than any satellite in our own Solar system. Such a simple restriction is therefore already suggestive of certain intrinsic properties of the majority of exomoons, given their non-detection so far.

In order to form an exomoon massive enough to be detectable, in accordance with the aforementioned criteria, it is required to have an unusually (by Solar system standards) large satellite-to-planet mass ratio, or else a very massive host planet. To illustrate the point, an exomoon around an Earth-analogue, requires a satellite to planet mass ratio of 1:5 in order to be detectable, twice than the Pluto-Charon mass ratio, which represents the largest mass ratio known in the Solar system. Typical in-situ formation of satellites inside cirumplanetary discs results in satellite-to-planet mass ratios of the order of $\sim 10^{-4}$ \citep{CanupWard-2006}, although more recent models \citep{CilibrasiEtAl-2018,InderbitziEtAl-2019} show that statistically, more massive satellites can dwell inside the rare tail in the mass distribution. It is therefore an unlikely way of forming currently-detectable exomoons, unless their host planets are in the super-Jupiter mass range (see also \cite{Heller-2014}, referring to the orbital sampling effect for a similar conclusion). In contrary, giant impacts are readily capable of forming satellites with large satellite-to-planet mass ratios. The most notable examples in the Solar system are the giant impact scenario that formed the Pluto-Charon system \citep{Canup-2005} and the one that formed the Earth-Moon system \citep{CanupAsphaug-2001}, with satellite-to-planet mass ratios of $\sim$$10^{-1}$ and $\sim$$10^{-2}$ respectively. Such collision geometries involving solid bodies are certainly plausible in the late stages of terrestrial planet formation \citep{ElserEtAl-2011,Chambers-2013}, and therefore might credibly give rise to massive exomoons around Earth-like or Super-Earth exoplanets.

Our goal in this paper is therefore to map the collision phase space relevant to the formation of massive exomoons around super terrestrial planets, including new formation pathways which have not yet been suggested in the existing collision formation literature, presented in Section \ref{S:Formation}. We then briefly introduce in Section \ref{S:Methods} the model used for hydrodynamical collision simulations, the considerations for our parameter space, and introduce our pre- and post-processing algorithms, in addition to our follow up N-body simulation setup. In Section \ref{S:Results} we present the simulation results, and discuss their implications in Section \ref{S:discussion}, including some predictions of exomoon detections around super-terrestrials, when using present-day or future instruments.

\section{Collisional formation of exomoons}\label{S:Formation}	
Most simulation studies of giant impacts have focused on the collisional phase space conductive to the formation of Solar system planets and satellites \citep{Barr-2016}. Despite an extensive collision simulation literature, there have only been a few studies that investigated hydrodynamical giant impact simulations relevant to exoplanets that are more massive than the Earth \citep{GendaAbe-2003,MarcusEtAl-2010a,MarcusEtAl-2010b,LiuEtAl-2015,InamdarSchlichting-2015,InamdarSchlichting-2016,BarrBruckSyal-2017,BierstekerSchlichting-2019}. In particular, only \citet{BarrBruckSyal-2017} (hereafter BB17) focus on the formation of exosolar satellites (or rather the discs from which they accreted), while all the rest examine the effects on the exoplanets themselves.
	 
The study of BB17 examines collisions onto rocky exoplanets up to 10 $M_\oplus$. Their goal is to identify the collision phase space capable of generating debris discs massive enough to form detectable exomoons by present-day technology. They use a Eulerian Adaptive Mesh Refinement (AMR) CTH shock physics code, and examine different masses, impact geometries and velocities. While BB17 manage to demonstrate, for the first time, detailed simulations capable of forming very massive proto-satellite discs -- only 2 cases in a suite of 28 impact simulations (7\%) result in a disc mass of $\sim$0.3 $M_\oplus$, hence beyond the 0.2 $M_\oplus$ criteria. Furthermore, the disc mass is only a hard upper limit on the mass of the exomoon that could form. The actual fraction of mass in the disc that coagulates into a moon is not immediately clear. It primarily depends on the initial specific angular momentum of the disc, and also on how one chooses to model the disc.
	
In N-body disc models that assume a particulate disc of condensed, solid particles, thus neglecting the presence of vapour, about 10--55\% of the mass of the disc would go into the satellite \citep{KokuboEtAl-2000}. In more complex hybrid models that consist of a fluid model for the disc inside the Roche limit and an N-body code to describe accretion outside the Roche limit, about 20-–45\% of the mass of the disc would go into the satellite \citep{SalmonCanup-2012}. The hybrid models provide a more realistic view since gravitational tidal forces both prevent accretion inside the Roche limit and also disrupt planet-bound eccentric clumps, scattered from outside the Roche limit by close encounters. A sufficiently energetic giant impact dictates that this zone is initially in a state of a two-phase liquid/vapour silicate 'foam' and as such its evolution is controlled by the balance between viscous heat dissipation (further inducing vaporization) and radiative cooling \citep{Ward-2012}. This disc spreads inward toward the planet, and outward beyond the edge of the Roche limit, where newly spawned clumps and their corresponding angular momentum join the particulate, satellite accreting disc. The complexity of such models, however, entails large uncertainties on the disc physics \citep{CharnozMichaut-2015}. The problem is more accentuated when one considers the conditions applicable to the formation of detectable exomoons around super-terrestrial planets, in which more massive or hotter discs are involved. Ignoring such complications, that is, if the aforementioned accretion studies are scalable to larger masses, it would suggest no more than about half the mass of the disc would form the final satellite. Given that assumption, the two outlier cases in the study of BB17 form exomoons in which the final mass is less than $\sim$0.15 $M_\oplus$, hence below the detecion criteria.
	
The primary goal of this study is therefore not only to reproduce, but also \emph{extend} the BB17 collision phase space in an attempt to identify more likely host planets or paths to forming massive exomoons, and thoroughly characterizing such moons in terms of their mass, composition and origin of materials. We also newly follow up on the long-term coagulation of satellites in the resulting SPH discs. In the following paragraphs we consider: (a) more giant impact scenarios which we speculate are likely to form massive discs; (b) the formation of massive exomoons through consecutive, multiple impacts, rather than in a single giant impact; and (c) a lower limit mass for the satellites which coagulate from the ensuing disc, using detailed N-body simulations.

\begin{figure}
	\begin{center}	
		\includegraphics[scale=0.77]{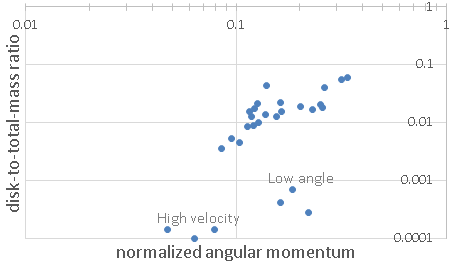}
		\caption{A reproduction of Figure 6 from BB17, plotting $M_\mathrm{o}/M_\mathrm{T}$ as a function of $J_\mathrm{col}$. 6 outlier cases in a suite of 28 collisions produce low mass discs and do not fit with the rest of the data, having either high impact velocities or low impact angles.}
		\label{fig:Barr_disc_mass}
	\end{center}	
\end{figure}

In order to include additional collision simulations that might increase the chance of forming massive exomoons, we look more closely at the BB17 collision phase space for specific hints. Figure \ref{fig:Barr_disc_mass} is a reproduction of Figure 6 in the study of BB17, and shows \emph{their} results for the disc-to-total mass ratio ($M_\mathrm{o}/M_\mathrm{T}$) as a function of the normalized angular momentum ($J_\mathrm{col}$). The latter is given by \citet{Canup-2005}:

\begin{equation}
J_\mathrm{col} = \sqrt{2}f(\gamma)\sin{\theta}\frac{v_\mathrm{imp}}{v_\mathrm{esc}}
\end{equation}

\noindent where $\theta$ is the impact angle, $v_\mathrm{imp}$ the impact velocity, $v_\mathrm{esc}$ the escape velocity, $\gamma$ the impactor-to-total mass ratio and $f(\gamma)=\gamma(1-\gamma)[\gamma^{1/3}+(1-\gamma)^{1/3}]^{1/2}$. As can be seen, $M_\mathrm{o}/M_\mathrm{T}$ and $J_\mathrm{col}$ are well-correlated, with the exception of six outlier cases that result in very low disc masses and do not fit the rest of the data. We have identified the outlier cases in the study of BB17 to be consecutively, \emph{velocity6} to \emph{angle3} in Table \ref{tab:suite}. The latter correspond to all the cases with either a high velocity or a low impact angle. In other words -- high velocity and low impact angle appear counter conductive to the formation of massive debris disc, and thus we judge them as incompatible avenues to forming massive exomoons.

In the upper part of Table \ref{tab:suite}, 19 simulations have approximately the canonical Lunar-forming $\gamma$ value ($\sim$0.11). The remaining 9 (in a total of 28 simulations) have much larger $\gamma$ values, and they form 7 of the 9 discs with the highest disc-to-total mass ratios. We thus judge large $\gamma$ to generate favourable outcomes, perhaps unsurprisingly, as they are more compatible with the Pluto-Charon impact scenario. In Section \ref{SS:ParameterSpace} we describe our additional simulations with a large value of $\gamma$.  

\begin{figure}
	\begin{center}	
		\includegraphics[scale=0.32]{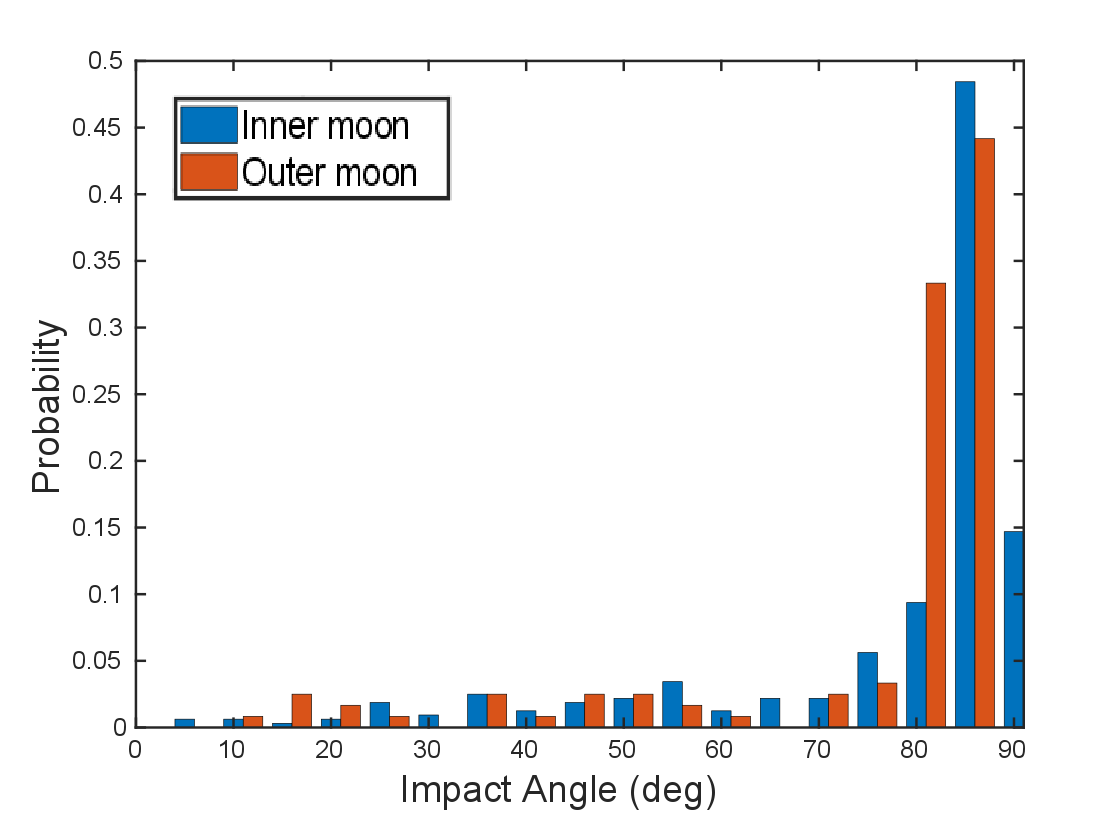}
		\caption{The distrbution of moonfall impact angles triggered by the gravitational perturbations between an inner and an outer moon ($\theta$=0\textdegree: head-on impact; $\theta$=90\textdegree: exteremely grazing impact) based on analysis of data from the \citet{CitronEtAl-2018} study.}
		\label{fig:impact-angles}
	\end{center}	
\end{figure}

As previously mentioned, we also consider the stepwise growth of exomoons through multiple, rather than a single impact. This sequence of impacts is simply part of the critical collisional evolution that naturally takes place during the last stages of terrestrial-type planet accretion. Multiple impacts create multiple exomoons. The tidal evolution and migration of these formed exomoons and their mutual gravitational perturbations \citep{RufuEtAl-2017,CitronEtAl-2018} could then result in several possible evolutionary outcomes, with roughly equal probabilities, including collisions among two exomoons (eventually growing into a more massive final exomoon), ejection of exomoons, or their recollisions with the host planet \citep{MalamudEtAl-2018}. Estimating the mass of such an exomoon, formed through mergers, is beyond the scope of this paper, since it requires a large set of N-body and moon-moon impact simulations. Assuming that such a moon can form, however, subsequent impacts onto the planet could in principal clear the moon by either ejecting it, or triggering a moonfall. For the latter case, we wish to investigate what would be the likely result. Based on the statistical analysis of \citet{CitronEtAl-2018}, most moonfalls have extremely grazing geometries, with $\theta\sim$90\textdegree, as shown in Figure \ref{fig:impact-angles}. It remains to be checked if extremely grazing moonfalls do not always result in the complete-loss of moon material, but instead give rise to a new generation of intact moons that retain most of their original mass (as was done by \citet{MalamudEtAl-2018} for Earth-sized planets). If this behaviour is shown here to be typical of super-terrestrial planets as well, then in the framework of multiple-impact formation, moonfalls may have a high probability of continuing the process of exomoon growth, rather than restarting it.

We note that in the course of this study, we will have an opportunity to perform a comparison between impact simulations using our Lagrangian SPH method, and the Eulerian AMR method in the BB17 study. This in itself has some important value, since (to our knowledge) only one other comparison study has been performed until now for planetary impacts \citep{CanupEtAl-2013}, and it involved only Lunar-forming scenarios. While a fully consistent comparison is difficult when the precise details involved (set-up procedures, pre-processing and post-processing algorithms) are handled by different groups (see e.g., Section \ref{SS:Caveats}), we can nevertheless show if using the SPH or AMR methods broadly result in the same overall trends in the data, and in particular the same amount of exomoons or discs which are potentially detectable using present-day instruments.

Finally, in order to estimate the lower-limit mass of the satellite/s that emerge from the disc, we hand over the resulting SPH discs to a more efficient, N-body code, for long-term tracking of the coagulation of satellites in the disc. The details involving our N-body and SPH simulations are given below in Section \ref{S:Methods}.

\section{Methods}\label{S:Methods}
	
\subsection{Hydrodynamical code outline}\label{SS:Outline}
We perform hydrodynamical collision simulations using an SPH code developed by \citet{SchaferEtAl-2016}. The code is implemented via CUDA, and runs on graphics processing units (GPU), with a substantial improvement in computation time, on the order of several $\sim10^{1}-10^{2}$ times faster for a single GPU compared to a single CPU, depending on the precise GPU architecture. The code has already been successfully applied to several studies involving hydrodynamical modeling \citep{DvorakEtAl-2015,MaindlEtAl-2015,HaghighipourEtAl-2016,WandelEtAl-2017,SchaferEtAl-2017,BurgerEtAl-2018,HaghighipourEtAl-2018,MalamudEtAl-2018,MalamudPerets-2019a,MalamudPerets-2019b}.
	
The code implements a Barnes-Hut tree that allows for treatment of self-gravity, as well as gas, fluid, elastic, and plastic solid bodies, including a failure model for brittle materials. Given the analysis of \citet{BurgerSchafer-2017} however, and the typical mass and velocity considered for our impactors and targets (see Section \ref{SS:ParameterSpace}), we perform our simulations in full hydrodynamic mode, i.e., neglecting solid-body physics, being less computationally expensive. We use the M-ANEOS equation of state (EOS), in compatibility with the BB17 study. Our M-ANEOS parameter input files are derived from \citet{Melosh-2007}.
	
\subsection{Collision parameter space}\label{SS:ParameterSpace}
We are exploring the phase space of potential large-exomoon forming impacts, which are more likely to be detectable, even with more advanced instruments than currently available.

As a starting point, we are repeating the impacts considered in the study of BB17, albeit using a different numerical method. All other parameters being equal, we can determine whether using an SPH code instead of the AMR code of BB17, might in itself improve or impede the formation of massive exomoons. We therefore analyse the BB17 suite of collisions, using identical composition and EOS. The BB17 suite of collisions is listed in the upper part of Table \ref{tab:suite}.	
	
As discussed in Section \ref{S:Formation}, we also extend the BB17 collision phase space. We identified giant-impact scenarios with a high impactor-to-total mass ratio, as being conductive to forming massive discs. Accordingly we set 4 new simulations similar to \emph{pc\_earth} and \emph{big\_earth} in the study of BB17, albeit with more massive planets, and also with equal mass impactor-target scenarios, as seen in the lower section of Table \ref{tab:suite} (\emph{bigger1-2} and \emph{biggest1-2}).
	
To complete our suite of simulations, we consider six additional moonfall cases, whereby existing exomoons with masses 0.05, 0.1, 0.2, 0.3, 0.4 and 0.5 $M_\oplus$ (\emph{moonfall1-6}) respectively, are gravitationally perturbed to collide with their host super-Earth planet at $\theta\sim$90\textdegree, as discussed in Section \ref{S:Formation} (in the context of multiple-impact exomoon growth). For simplicity we assume that their impact velocity equals the mutual escape velocity $v_\mathrm{esc}$, being a reasonable upper limit, according to the analysis of the impact velocity distribution data from the \citep{CitronEtAl-2018} study.

\subsection{Initial setup}\label{SS:Initial}
We consider differentiated impactors and targets composed of 30\% iron and 70\% dunite by mass. Both impactors and targets are non-rotating prior to impact, and are initially placed at touching distance. Including rotation as a free parameter would entail a very large increase in the number of simulations, especially if one assumes a non-zero angle between the collisional and equatorial planes. Here we assume no rotation and a coplanar collision geometry, in order to comply with the BB17 study. As we discuss in Section \ref{SS:Results_TidalStability}, the initial rotation in particular could have some significance, and we suggest to explore it in future dedicated studies (see Section \ref{SS:Future}).

The initial setup of each simulation is calculated via a pre-processing step, in which both impactor and target are generated with relaxed internal structures, i.e. having hydrostatic density profiles and internal energy values from adiabatic compression, following the algorithm provided in appendix A of \citet{BurgerEtAl-2018}. This self-consistent semi-analytical calculation (i.e., using the same constituent physical relations as in the SPH model) equivalently replaces the otherwise necessary and far slower process of simulating each body in isolation for several hours, letting its particles settle into a hydrostatically equilibrated state prior to the collision (as done e.g., in the work of \citet{CanupEtAl-2013} or \citet{SchaferEtAl-2016}).
		
Altogether we have 38 simulations. The simulations were performed on the bwForCluster BinAC, at T{\"u}bingen University. The GPU model used is NVIDIA Tesla K80. Each simulation ran on a single dedicated GPU for approximately 7-10 days on average, tracking the first 28 hours post collision to comply with BB17 on what they considered a complete simulation. In some simulations (such as $pc\_earth$ and $velocity5$) we will triple the simulation duration for accurate outcomes. In total we thus have $\sim$60 weeks of GPU time.

We perform our simulations using a high resolution of $10^{6}$ SPH particles, resulting in sensible and practical runtimes, as mentioned above. We nevertheless caution that the SPH method has well-known issues arising in low-density regions (see e.g. \cite{ReinhardtStadel-2017}), and that even higher resolution simulations should be performed in the future, to corroborate our results. 
	
\subsection{Debris-disc mass analysis}\label{SS:DiskMass}
Planet scale impact simulations often result in a cloud of particles, some of which will accrete onto the planet, other remain in a bound proto-satellite disc, and some escape the system. In order to determine which particles go where, an analysis is performed as a post-processing step. Our algorithm resembles the one used by BB17. BB17 refer to the procedures described by \cite{CanupEtAl-2001} as the basis for their analysis. Given the latter, we note that some of the details required in order to compare our approaches are missing or are incompatible with our interpretation of the data (explanation will be provided below). As a consequence, we point out that there may be minor variations in our respective analyses results, however we judge them to have a small effect because the overall scheme follows essentially a similar classification approach. 

Our detailed classification algorithm follows this 5-step algorithm:

(a) We find physical fragments (clumps) of spatially connected SPH particles using a friends-of-friends algorithm. 

(b) The fragments are then sorted in descending order according to their mass.

(c) We classify these fragments in two categories: gravitationally bound (GB) to the planet and gravitationally unbound (GUB). The first fragment (i.e., the most massive, in this case the target/proto-planet) is initially the only one marked as GB, and the rest are marked as GUB. 

(d) We calculate: 
\begin{equation}
\vec{r}_\mathrm{GB}=\frac{\sum_{j}m_{j}\vec{r}_{j}}{\sum_{j}m_{j}},\vec{v}_\mathrm{GB}=\frac{\sum_{j}m_{j}\vec{v}_{j}}{\sum_{j}m_{j}}
\label{eq:CenterOfMass}
\end{equation}
where $\vec{r}_\mathrm{GB}$ and $\vec{v}_\mathrm{GB}$ are the center of mass position and velocity of GB fragments, $j$ denoting indices of GB fragments and $m_{j}$, $\vec{r}_{j}$ and $\vec{v}_{j}$ are the corresponding mass, position and velocity of each fragment.

Then, for each fragment marked as GUB we check if the kinetic energy is lower than the potential energy: 
\begin{equation}
\frac{\lvert\vec{v}_{i}-\vec{v}_\mathrm{GB}\rvert^{2}}{2}<\frac{G\left(M_\mathrm{GB}+m_{i}\right)}{\lvert\vec{r}_{i}-\vec{r}_\mathrm{GB}\rvert}
\label{eq:Bound}
\end{equation}
where $G$ is the gravitational constant and $M_\mathrm{GB}$ is the summed mass of gravitationally bound fragments. $m_{i}$, $\vec{r}_{i}$ and $\vec{v}_{i}$ are the fragment mass, position and velocity, $i$ denoting indices of GUB fragments. If equation \ref{eq:Bound} is satisfied then fragment $i$ is switched from GUB to GB. We iterate on step (d) until convergence (no change in fragment classification within an iteration) is achieved. At this point we deem the mass of the planet $M_\mathrm{P}$ as equal to that of most massive GB fragment, while the other GB fragments will be classified as either belonging to the planet or to the bound disc according to the following, final step.

(e) By the following steps, we calculate the fragment pericentre $q_{j}$, $j$ denoting the indices of GB fragments, however excluding P (planet bound) fragments, i.e., deemed as belonging to the planet and contributing to its mass. We first define the fragment relative position and velocity vectors:

\begin{equation}
\mathbf{\vec{r}}_{j}=(\vec{r}_{j}-\vec{r}_\mathrm{P}),\mathbf{\vec{v}}_{j}=(\vec{v}_{j}-\vec{v}_\mathrm{P})
\label{eq:RelativeVectors}
\end{equation}

Where $\vec{r}_\mathrm{P}$ and $\vec{v}_\mathrm{P}$ are the planet/planet-bound (P) fragments' centre of mass coordinates and are calculated in a similar way to equation \ref{eq:CenterOfMass}. Then the fragment specific energy is calculated from:

\begin{equation}
E_{j}=\frac{{\lvert\mathbf{\vec{v}}_{j}\rvert}^{2}}{2}-\frac{G(M_\mathrm{P}+m_{j})}{\lvert\mathbf{\vec{r}}_{j}\rvert}
\label{eq:Energy}
\end{equation}

Based on the specific orbital energy, we calculate the fragment semi-major axes from the relation $a_{j}=-G(M_\mathrm{P}+m_{j})/2E_{j}$. The eccentricity vector is given by:

\begin{equation}
\vec{e}_{j}=\left( \frac{{\lvert\mathbf{\vec{v}}_{j}\rvert}^{2}}{G(M_\mathrm{P}+m_{j})} - \frac{1}{\lvert\mathbf{\vec{r}}_{j}\rvert} \right) \mathbf{\vec{r}}_{j} - \frac{\mathbf{\vec{r}}_{j} \cdot \mathbf{\vec{v}}_{j}}{G(M_\mathrm{P}+m_{j})} \mathbf{\vec{v}}_{j}
\label{eq:Eccentricity}
\end{equation}

Finally we obtain the pericentre from $q_{j}=a_{j}(1-e_{j})$. Since we assume that the collision geometry is coplanar to the planet's equator, each fragment that satisfies $q_{j}<R_\mathrm{P\_equ}$, i.e., has its pericentre residing inside the planet's physical cross-section $R_\mathrm{P\_equ}$ -- may be classified as P, having a planet-bound trajectory. We iterate on step (e) until convergence is achieved. The calculation of $R_\mathrm{P\_equ}$ is specified in Section \ref{SS:Stability}.

Once the algorithm is complete we have the unbound mass $M_\mathrm{U}=M_\mathrm{GUB}$, planet-bound mass $M_\mathrm{P}$ and disc mass $M_\mathrm{D}=(M_\mathrm{GB}-M_\mathrm{P})$, respectively. We also track for each one, their respective compositions and the origin of SPH particles (impactor or target). 

\subsection{Tidal stability analysis}\label{SS:Stability}
In addition to calculating the mass of the disc (to check its potential of forming a massive exomoon), we also discuss if a satellite can survive its post-accretion tidal evolution. The Roche radius $r_\mathrm{roche}$ is the distance below which tidal forces would break the satellite apart, and therefore the initial accretion distance of the satellite has to be greater. The Roche radius is given by $r_\mathrm{roche}=2.44R_\mathrm{P}(\rho_\mathrm{P}/\rho_\mathrm{S})^{1/3}$, where $R_\mathrm{P}$ is the planet's radius, $\rho_\mathrm{P}$ the planet's density and $\rho_S$ the satellite's density. The mean distance of satellite formation in giant-impact simulations is slightly beyond the Roche radius, at around 1.3$r_\mathrm{roche}$ \citep{ElserEtAl-2011}.

After the initial accretion, the satellite's orbit evolves by tidal interactions with the planet, depending on its position relative to the synchronous radius $r_\mathrm{sync}$, the distance at which the satellite's circular orbital period equals the rotation period of the planet, or in other words, the satellite's orbital mean motion ($n$) equals the initial rotation rate of the planet ($\Omega$). The synchronous radius $r_\mathrm{sync}$ depends on the rotation rate of the planet, and can be obtained by equating the gravitational acceleration ($G M_\mathrm{P} / r^2$) and the centripetal acceleration ($v^2/r$ or $\Omega^2 r$), giving $r_\mathrm{sync} = [(GM_\mathrm{P})/(\Omega^2)]^{1/3}$.

Satellites that form inside the synchronous radius, orbit their planets faster than their planet rotates. In this case the tidal bulge they raise on the planet lags behind, and acts to decelerate them, and so they spiral towards the planet and are ultimately destroyed when they cross the Roche radius. However, satellites that form outside the synchronous radius, recede from the planet as angular momentum is tidally transferred from the planet to the satellite. In either case, more massive satellites tidally evolve faster. One can reach a conclusion on the tidal fate of a satellite, comparing the two radii. Since in giant-impacts a satellite typically forms from a disc just outside the Roche radius, it implies that if $r_\mathrm{sync}<$ $\sim$$1.3r_\mathrm{roche}$, the satellite evolves outwards and survives. It is also understood that in slowly rotating planets, $r_\mathrm{sync}$ moves outwards, therefore making it much more difficult to form tidally-stable satellites. We note that the aforementioned discussion applies to prograde satellites. Retrograde satellites which orbit in the opposite sense relative to the planet's rotation, will of course also in-spiral. In this paper however, we start all our simulations with non-rotating targets and impactors, hence all satellites will orbit in the same sense.

To complete the set of equations, the initial rotation rate of the planet $\Omega$ can be calculated from the angular momentum of the material judged to constitute its mass. For this calculation we take only the mass of the largest fragment, labelled $M_\mathrm{1}$. We note that by the end of the simulation, the small fraction of planet-bound material which has not yet accreted, cannot contribute more than a few \% to the final planet mass anyway, and $\Omega$ is not expected to change significantly. For $M_\mathrm{1}$ we then find the total angular momentum $\vec{L}=\sum m_{par}\left(\vec{r}_{par}\times\vec{v}_{par}\right)$ by the summation of its individual particle angular momenta, where $m_{par}$ denotes SPH particle mass and $\vec{r}_{par}$ and $\vec{v}_{par}$ denote SPH particle relative (to fragment's center of mass) position and velocity. We then calculate the angular momentum unit vector $\hat{L}=\vec{L}/\lvert\vec{L}\rvert$, which points towards the direction of the rotation axis. In order to get the rotation rate $\Omega$ we calculate $\vec{R}=\hat{L}\times\vec{r}_{par}$, the distance vector from the rotation axis to the particle relative position. Then the total moment of inertia is similarly given by $I=\sum m_{par}{\lvert\vec{R}\rvert}^{2}$, and the rotation rate by $\Omega=I/\lvert\vec{L}\rvert$.

For a rotating object, we expect to have some degree of flattening $f$, such that:
\begin{equation}
f=\frac{R_\mathrm{equ}-R_\mathrm{pol}}{R_\mathrm{equ}}
\label{eq:Flattening}
\end{equation}

where $R_\mathrm{equ}$ and $R_\mathrm{pol}$ are its equatorial and polar radii respectively. For such an object with mass $M$, the density $\rho$ is generally given by:
\begin{equation}
\rho=\frac{3M}{4\pi {R_\mathrm{equ}}^2 R_\mathrm{pol}}
\label{eq:PlanetDensity}
\end{equation}

In order to calculate the planet's density $\rho_\mathrm{P}$ we thus need to know its equatorial and polar radii. However the latter two quantities cannot be calculated directly before all planet-bound material has accreted, so we will assume for simplicity that $\rho_\mathrm{P}$ equals the density of the largest fragment, hereafter labelled $\rho_\mathrm{1}$. Such an assumption is judicious since the mass of the planet is, as previously mentioned, very close to the mass of the proto-planet (the first) fragment at the end of our simulations, while non-negligible changes to the density typically entail a large change in the mass (hence the pressure by self-gravity). By the same consideration, and since changes in $\Omega$ were also regarded as negligible, we will assume consistently that the flattening $f_\mathrm{P}=f_\mathrm{1}$. Now $\rho_\mathrm{1}$ can be directly calculated from Equation \ref{eq:PlanetDensity} substituting $M$ for $M_\mathrm{1}$. The equatorial and polar radii of the biggest fragment are physically (directly) computed by considering its constituent particles, such that $R_\mathrm{1\_equ}=max(\lvert\vec{R}\rvert)$ and $R_\mathrm{1\_pol}=max(\lvert\vec{r}\rvert), \forall \lvert\vec{R}\rvert <\sim$sml (the smoothing length distance). From the latter we can calculate $f_\mathrm{1}$ using Equation \ref{eq:Flattening}. We note that these equatorial and polar radii values, which are measured relative to the planet rotation axis, are simply indicative of the planet's shape, and therefore not always identical to those of a hydrostatically flattened ellipsoid (see Figure \ref{fig:gamma3Profile} and associated discussion). In that sense $f_\mathrm{P}$ is simply the ratio given by Equation \ref{eq:Flattening}, which helps us calculate the planet's physical cross-section.

In contrast to the planet, the satellite has not fully accreted at the end of the simulation, because the timescale for full accretion is at least two orders of magnitude longer. We thus have no direct information about its physical properties, but we do have information about the disc from which it will coagulate. In order to calculate $\rho_\mathrm{S}$ we will use the fact that the satellite is not nearly as massive as the planet, and therefore as a heuristic approach, it could be more readily estimated from $1/\sum_k (X_k/\varrho_k)$, $k$ denoting the indices of the disc's constituent materials, $X_k$ the relative material mass fractions and $\varrho_k$ their corresponding specific densities. We note that unlike in Solar system satellites, for extremely massive exomoons we expect the actual $\rho_\mathrm{S}$ to be somewhat larger than this estimation, and therefore our ensuing $r_\mathrm{roche}$ should be considered an upper limit.

Finally, we rewrite Equations \ref{eq:Flattening} and \ref{eq:PlanetDensity} for $M_\mathrm{P}$ and $\rho_\mathrm{P}$:

\begin{equation}
R_\mathrm{P\_equ}=\left(\frac{3M_\mathrm{P}}{4\pi \rho_\mathrm{P}(1-f_\mathrm{P})}\right)^{1/3}
\label{eq:PlanetEquR}
\end{equation}

\begin{equation}
R_\mathrm{P\_pol}=R_\mathrm{P\_equ}(1-f_\mathrm{P})
\label{eq:PlanetPolR}
\end{equation}

The effective planet radius $R_\mathrm{P}$ is then:
\begin{equation}
R_\mathrm{P}=\left({R_\mathrm{P\_equ}}^2 R_\mathrm{P\_pol}\right)^{1/3}
\label{eq:EffectiveR}
\end{equation}

Equations \ref{eq:PlanetEquR}-\ref{eq:EffectiveR} are iteratively re-evaluated in step (e) of Section \ref{SS:DiskMass}, in order to obtain the disc mass.

\subsection{Algorithm differences from previous studies}\label{SS:Caveats}
Since we will be comparing our results with previous studies, we wish to note as a caveat that the technique for calculating the equatorial radius in the BB17 study, and therefore the planet and disc masses, is based on an algorithm from an Earth-related study by \cite{CanupEtAl-2001}, wherein an iterative process is used to estimate these radii from the Earth's mass and density, assuming the latter is equal to the known present-day Earth density value. This value is of course unsuitable for the much more massive exo-planets considered by BB17, and would lead to a 'too-small' planet radius, and therefore Roche radius, which also bears directly on the tidal stability conclusion. BB17 however did not specify what density value they did use in their study, nor the details of how it was calculated. Our calculation of the planet's density is however based on measuring it directly from its physical size and mass, as was described above.

\begin{figure}
	\begin{center}	
		\includegraphics[scale=0.326]{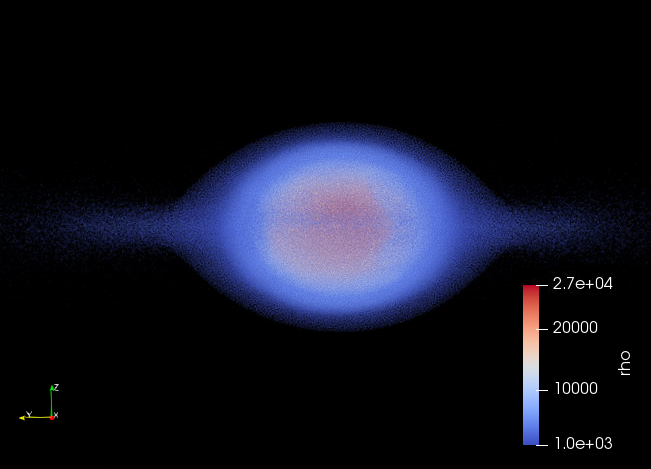}
		\caption{Edge-on view of the planet at the end of the \emph{gamma3} simulation. The (semi-transparent) colour scheme shows the planet's density in kg $\times$ m$^{-3}$. The inner planet is surrounded by a cloud of relatively high-density (>1000 kg $\times$ m$^{-3}$), and extremely hot silicate material (up to 40000 K). The planet's collision cross-section does not comply with a standard flattened ellipsoid shape.}
		\label{fig:gamma3Profile}
	\end{center}	
\end{figure}

Another difference in the algorithm is in the calculation of the planet's flattening. \cite{CanupEtAl-2001} assume, as we do, that during the initial disc formation process, only a clump of material whose pericentre resides inside the equatorial radius, will merge with the planet (see step (e) of Section \ref{SS:DiskMass}). However, their approach is to calculate this physical cross-section from the planet's flattening coefficient, which is in turn calculated according to a prescribed formula \citep{Kaula-1968}. The latter is given as a function of the planet's rotation rate, which can be calculated from its angular momentum. Our inspection of the data, however, shows that for extremely energetic impacts the actual planet's post-collision shape is not always that of a standard flattened ellipsoid in hydrostatic equilibrium (unlike in canonical Lunar-forming giant impacts). In some cases, we see an extended region of relatively high density material (>1000 kg $\times$ m$^{-3}$) at temperatures of several $\sim$10000 K, as in Figure \ref{fig:gamma3Profile}. Previous studies argue that the relevant viscous timescale of such material is much longer than the initial disc formation timescale, dominated by gravity \citep{Ward-2012}, hence the effective cross-section for clump mergers can be larger relative to the one obtained by using the prescribed, rotation-dependent flattening coefficient formula from \cite{CanupEtAl-2001}. We therefore define the flattening coefficient according to Equation \ref{eq:Flattening}, where the planet's equatorial and polar radii are measured with respect to its data-derived shape, as previously described.

Given the arguments above, we caution that there may be some differences primarily between our planet radius calculation and that of BB17. Therefore the Roche radius and to a lesser extent also the disc mass analysis, may result in somewhat different values, despite using mostly identical equations and procedures.

We also point out that a few of our collisions do not necessarily form a disc of debris, but rather result in graze \& merge/capture scenarios, which form intact moons. In that case, the Roche radius is unrelated to tidal stability because it is not necessarily indicative of the initial distance of the satellite. It is rather the satellite's pericentre that we must compare to $r_\mathrm{sync}$, in order to gain some approximative understanding of how the orbit will develop. Such considerations will be discussed in Section \ref{S:Results}.

\subsection{N-body code outline}\label{SS:NBody_Outline}
In order to study the coagulation of exomoons we introduce a new N-body follow-up calculation. Some of the discs generated in Table \ref{tab:suite} are insufficiently massive. These discs have relatively few particles, and are not expected to form significant exomoons. Since our focus in this study is to form massive exomoons, we arbitrarily select 0.1$M_\oplus$ as the limiting disc mass, above which we will perform N-body follow-ups. According to this selection criteria, we have 12 discs for which we simulate the coagulation of exomoons. Other discs are ignored.

The discs ensuing from SPH simulations are handed over to the open-source N-body code REBOUND, via a special tool which we have developed. The hand-off tool initially reads our SPH output files which are in turn synthesized based on an analysis that finds physical fragments (clumps) of spatially connected SPH particles using a friends-of-friends algorithm. Fragment data is then passed on as recognizable input particles, to a modified REBOUND code. The N-body setup is designed to keep a detailed record of mergers, which are treated as perfect mergers, using the existing REBOUND reb\_collision\_resolve mechanism. We have modified the REBOUND source code to keep track of the relative compositions of merged particles (utilizing the 'additional properties' built-in formalism), which we then use in order to calculate a more realistic physical collision-radius for the REBOUND particles. Additionally, by the same formalism we also keep track of the material origin, i.e., if it came from the impactor or the target. Finally, we modify the code to remove particles that, through close encounters, obtain hyperbolic trajectories during the simulation. We note that in principle, \cite{GrishinEtAl-2017} show that particles with apocentres outside $\sim 40\%$ the planet's Hill radius, should likewise be removed from the simulation. However, without prior assumptions on the host-star's mass, or the planet's orbit/inclination, there is insufficient information to constrain the Hill sphere or the exact stability criteria. We refrain from introducing any more free parameter to the simulation, thereby ignoring the potential influence of the host star.

The planet radius is set to match $r_\mathrm{roche}$ from Section \ref{SS:Stability}. We assume that circumplanetary material with inner-Roche trajectories would disrupt, and eventually end up inside the planet. Our assumption is however too restrictive. Complex models show that material in this inner disc zone would actually spread both inward toward the planet, and also outward beyond the edge of the Roche limit, where newly spawned clumps and their corresponding angular momentum join the particulate, satellite accreting disc (see Section \ref{S:Intro}). Hence, our follow-up simulations necessarily provide us with a \emph{lower limit} mass estimate for the satellite/s that eventually form.

For the integration we use IAS15 - a non-symplectic, fast, high-order integrator with adaptive time-stepping, accurate to machine precision over a billion orbits \citep{ReinSpiegel-2015}. Our implementation also utilizes openmp, to get about a 30\% improvement in runtime when using 8 cores in parallel (more cores gain no further improvement). Our integration time is set to 35 years. It is selected based on our observation that the mass, composition and material-origin of the major formed fragments converge after a few years of integration. Any longer-term dynamical evolution requires effective treatment/incorporation of tidal migration \citep{CitronEtAl-2018}, which begins to have comparable timescales and cannot be ignored, and also adding the star to the simulation, on top of the planet. However, as previously mentioned the complexity of our models is already significant and we do not wish to introduce any more free parameters for planet and satellites (e.g. the moment of inertia factors, love numbers, planet orbit, star mass, etc.). We thus ignore tidal migration or perturbations from the star, and truncate the simulation at the 35 year limit.

Altogether we have 12 simulations. The resolution is variable and depends on the outcome of our various SPH simulations, but is generally on the order of $10^{4}$ REBOUND particles (each representing a fragment, so they are not equal mass). The simulations were performed on the Astrophysics (Astro) iCore HPC cluster, at the Hebrew University of Jerusalem.

\section{Results}\label{S:Results}
The results from all simulations are summarized in Table \ref{tab:suite}, which shows, from left to right, the parameters for different scenarios (upper table, reproduction of the BB17 suite of collisions and lower table, new collisions in this study); the disc mass in the BB17 study; the disc mass + synchronous \& Roche radii + composition in this study; and the emerging exomoon mass and composition.

\begin{table*}
	\caption{Suite of collisions.}
	\begin{tabular}{*{14}{l|}}
		\hline
		\multicolumn{5}{|l|}{\bf Parameters} \vline & \multicolumn{1}{|l|}{\bf BB17} 		 \vline & \multicolumn{5}{|l|}{\bf SPH (this study)} \vline & \multicolumn{3}{|l|}{\bf N-body follow up}\\
		\hline
		{Name} & {$\frac{M_\mathrm{T}}{M_\mathrm{\oplus}}$} & {$\gamma$} & {$\frac{v_\mathrm{imp}}{v_\mathrm{esc}}$} & {$\theta$} & {$\frac{M_\mathrm{o}}{M_\mathrm{\oplus}}$} & {$\frac{M_\mathrm{o}}{M_\mathrm{\oplus}}$} & {$\frac{r_\mathrm{sync}}{10^4km}$} & {$\frac{r_\mathrm{roche}}{10^4km}$} & {$f_\mathrm{iron}$} & {$f_\mathrm{targ}$} & {$\frac{M_\mathrm{S}}{M_\mathrm{\oplus}}$} & {$f_\mathrm{iron}$}& {$f_\mathrm{targ}$}\\
		\hline
		big\_earth  & 2.311  & 0.325 & 0.929 & 62.13 & 0.127  & 0.069  & 1.162 & 2.086 & 0.587 & 0.036\\
		pc\_earth   & 1.013  & 0.327 & 0.983 & 61.92 & 0.061  & 0.083  & 0.966 & 1.473 & 0.767 & 0.042\\
		rocky\_exo2 & 0.266  & 0.146 & 0.986 & 50.55 & 0.006  & 0.004  & 0.805 & 1.066  & 0.437 & 0.174\\
		rocky\_exo3 & 0.466  & 0.142 & 0.962 & 50.55 & 0.006  & 0.007  & 0.938 & 1.298  & 0.424 & 0.213\\
		rocky\_exo4 & 2.324  & 0.123 & 0.862 & 50.55 & 0.04   & 0.036  & 1.669 & 2.188 & 0.459 & 0.294\\
		rocky\_exo5 & 3.805  & 0.119 & 0.825 & 50.55 & 0.032  & 0.051  & 1.916 & 2.576 & 0.467  & 0.266\\
		rocky\_exo6 & 7.208  & 0.115 & 0.774 & 50.55 & 0.033  & 0.027   & 2.165 & 3.269 & 0.381  & 0.253\\
		rocky\_exo7 & 11.151 & 0.111 & 0.736 & 50.55 & 0.058  & 0.007  & 2.439 & 3.872 & 0.286 & 0.289\\
		rocky\_exo8 & 18.066 & 0.106 & 0.691 & 50.55 & 0.063  & 0.001  & 2.851 & 4.83  & 0	    & 0.333\\
		ser119      & 1.015  & 0.130 & 0.922 & 50.55 & 0.014  & 0.008  & 1.246 & 1.684   & 0.408 & 0.317\\
		velocity3   & 7.208  & 0.115 & 0.864 & 50.55 & 0.112  & 0.104  & 2.306 & 3.203 & 0.444  & 0.257&0.016 & 0.241 & 0.349\\
		velocity4   & 7.208  & 0.115 & 0.953 & 50.55 & 0.071  & 0.057  & 2.252 & 3.201 & 0.444 & 0.282\\
		velocity5   & 7.208  & 0.115 & 1.042 & 50.55 &\bf0.311& 0.078  & 2.256 & 3.086 & 0.574 & 0.229\\
		velocity6   & 7.208  & 0.115 & 1.216 & 50.55 & 0.003  & 0.003  & 4.341 & 3.435 & 0.012 & 0.592\\
		velocity7   & 7.208  & 0.115 & 1.390 & 50.55 & 0.005  & 0.001  & 5.27  & 3.406 & 0.036 & 0.266\\
		velocity8   & 7.208  & 0.115 & 1.647 & 50.55 & 0.002  & 0	   & 6.617 & 3.42  & 0   & 0.258\\
		angle1      & 7.208  & 0.115 & 0.774 & 20.70 & 0.001  & 0      & 3.438 & 3.555 & 0	    & 0.668\\
		angle2      & 7.208  & 0.115 & 0.774 & 28.11 & 0	  & 0      & 2.875 & 3.554 & 0   & 0\\
		angle3      & 7.208  & 0.115 & 0.774 & 36.09 & 0.001  & 0      & 2.503 & 3.554 & 0   & 0.386\\
		angle4		& 7.208  & 0.115 & 0.774 & 62.07 & 0.09	  & 0.155  & 2.322 & 3.178 & 0.468 & 0.197& 0.039 & 0.22  & 0.286\\
		angle5      & 7.208  & 0.115 & 0.774 & 70.46 & 0.151  & 0.164  & 2.278 & 3.162 & 0.486 & 0.283& 0.064 & 0.25  & 0.009\\
		gamma1      & 7.126  & 0.196 & 0.778 & 50.55 & 0.112  & 0.092  & 1.775 & 3.303 & 0.309  & 0.298\\
		gamma2      & 7.131  & 0.321 & 0.778 & 50.55 & 0.12	  & 0.11   & 1.57  & 3.427 & 0.148 & 0.455& 0.005 & 0.101 & 0.588\\
		gamma3      & 7.264  & 0.451 & 0.771 & 50.55 &\bf0.298& 0.138  & 1.563 & 3.403 & 0.196 & 0.559& 0.052 & 0.188 & 0.582\\
		gamma4      & 7.518  & 0.442 & 0.757 & 50.55 & 0.136  & 0.135  & 1.572 & 3.418 & 0.23  & 0.541& 0.045 & 0.26  & 0.581\\
		gamma5      & 7.176  & 0.137 & 0.775 & 50.55 & 0.063  & 0.052  & 2.025 & 3.283 & 0.353 & 0.293\\
		gamma6      & 7.112  & 0.258 & 0.779 & 50.55 & 0.135  & 0.113  & 1.644 & 3.407 & 0.163 & 0.333& 0.002 & 0.02  & 0.55\\
		gamma7      & 7.182  & 0.386 & 0.775 & 50.55 & 0.148  & 0.139  & 1.561 & 3.39  & 0.197 & 0.466& 0.014 & 0.066 & 0.472\\
		\hline
		bigger1     & 7.2	 & 0.327 & 0.983 & 61.92 &		  &\bf0.298& 1.565 & 3.015 & 0.614 & 0.081& 0.077 & 0.317 & 0.144\\
		bigger2     & 7.2    & 0.5   & 0.983 & 61.92 &		  & 0.189  & 1.719 & 3.486 & 0.055 & 0.486& 0.186 & 0.55  & 0.486\\
		biggest1    & 11	 & 0.327 & 0.983 & 61.92 &		  &\bf0.256& 1.715 & 3.926 & 0.143 & 0.222& 0.104 & 0.069 & 0.178\\
		biggest2	& 11	 & 0.5   & 0.983 & 61.92 & 		  & 0.159  & 1.975 & 4.041 & 0.024 & 0.51 & 0.144 & 0.024 & 0.51\\
		moonfall1   & 7.2  	 & 0.007 & 1 	 & 89	 & 		  & 0.005  & 42.98 & 3.012 & 0.674 & 0.001\\
		moonfall2   & 7.2    & 0.014 & 1 	 & 89	 & 		  & 0.007  & 30.68 & 3.008 & 0.671 & 0\\
		moonfall3	& 7.2    & 0.028 & 1	 & 89	 & 		  & 0.186* & 23.42 & 2.937 & 0.729 & 0\\
		moonfall4   & 7.2  	 & 0.042 & 1 	 & 89	 & 		  & 0.235* & 15.93 & 2.906 & 0.753 & 0\\
		moonfall5   & 7.2    & 0.056 & 1 	 & 89	 & 		  & 0.326* & 14.94 & 2.893 & 0.753 & 0.705\\
		moonfall6	& 7.2    & 0.069 & 1	 & 89	 & 		  & 0.428* & 14.52 & 2.885 & 0.747 & 0\\
		\hline
	\end{tabular}
	\label{tab:suite}
	
	\begin{flushleft}
	>The upper part of the table consists of the 28 collisions performed by BB17 and repeated in this study. The lower part of the table lists the extended suite of collision simulations considered \emph{only} in this study.\\
	
	>Columns from left to right (notation from Section \ref{SS:ParameterSpace}): (1) simulation name ; (2) total mass of target and impactor in Earth mass units ; (3) impactor-to-target mass ratio ; (4) impact velocity in units of the mutual escape velocity ; (5) impact angle ; \textbf{AMR results from BB17:} (6) disc mass in Earth mass units ; \textbf{results from this study:} (7) disc mass in Earth mass units ; (8) synchronous rotation radius ; (9) Roche radius ; (10) disc iron fraction ; (11) disc target-material fraction. \textbf{SPH results from this study:} (7) disc mass in Earth mass units ; (8) synchronous rotation radius ; (9) Roche radius ; (10) disc iron fraction ; (11) disc target-material fraction. \textbf{follow-up N-body results from this study:} (12) mass of formed exomoon in Earth mass units ; (13) exomoon iron fraction ; (14) exomoon target-material fraction.\\
	
	>Highlighted values mark cases in which the disc mass exceeds the 0.2 $M_\oplus$ detection criteria.\\
	
	>Disc masses marked with * indicate intact satellites with pericentres well below the Roche limits. On second approach they will re-disrupt.
	\end{flushleft}
\end{table*}

\subsection{Disc masses}\label{SS:Results_DiscMass}
\begin{figure}
	\subfigure[0.17 hours]{\label{fig:exomoon1}\includegraphics[scale=0.26]{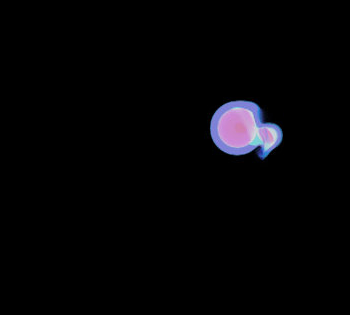}}
	\subfigure[1.5 hours]{\label{fig:exomoon2}\includegraphics[scale=0.26]{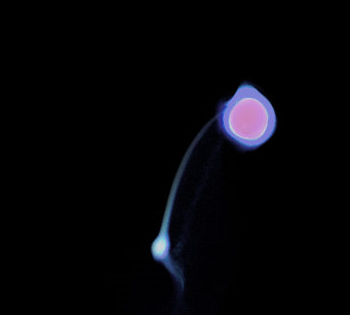}}
	\subfigure[3 hours]{\label{fig:exomoon3}\includegraphics[scale=0.325]{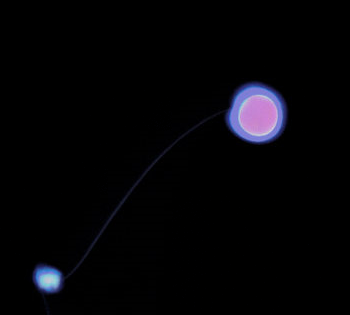}}
	\subfigure[34 hours]{\label{fig:exomoon4}\includegraphics[scale=0.26]{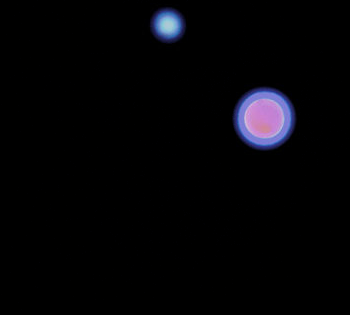}}
	\subfigure[35 hours]{\label{fig:exomoon5}\includegraphics[scale=0.26]{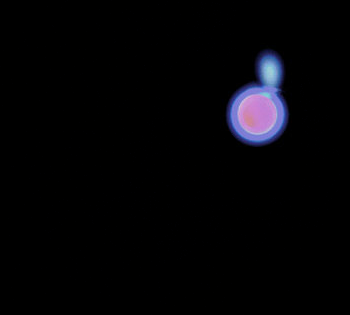}}
	\subfigure[36 hours]{\label{fig:exomoon6}\includegraphics[scale=0.26]{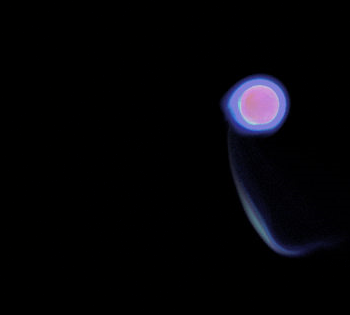}}
	\centering\subfigure{\includegraphics[scale=0.75]{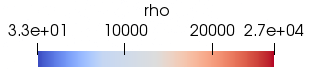}}
	\caption[r]{The \emph{velocity5} collision scenario: an original 0.83 M$_\mathrm{\oplus}$ impactor forms a massive $\sim$0.5 $M_\oplus$ exomoon (first 3 hours, Panels (a)-(c)), however during its subsequent close approach (Panels (d)-(f)) the exomoon disrupts inside the planet's tidal sphere, resulting in a graze \& merge scenario with relatively little debris. The (semi-transparent) colour scheme shows density in kg $\times$ m$^{-3}$ (see legend below). Resolution is $10^6$ SPH particles.}
	\label{fig:velocity5}
\end{figure}

Results from this study show that none of the discs from the original BB17 suite of 28 collisions, have a mass larger than the 0.2 $M_\oplus$ detection criteria. A close inspection of the data however, shows two cases, \emph{pc\_earth} and \emph{velocity5}, that stand out as unique graze \& capture/merger scenarios, respectively. Both cases generate intact exomoons, rather than discs. The former exomoon has a pericentre distance marginally outside the planet Roche limit ($1.54\times10^4$ km vs. $1.47\times10^4$ km), whereas the latter has a pericentre distance well inside the Roche limit. To investigate the detailed outcome of these cases we triple the fiducial simulation duration (Section \ref{SS:Outline}) to 90 h and follow the consequent re-entries of these exomoons, confirming that they both re-disrupt. As expected, the \emph{velocity5} initial exomoon disrupts violently, resulting in most of the mass entering the planet, while also leaving a considerable yet much lower fraction of its mass in the disc. Figure \ref{fig:velocity5} shows the initial phase of intact exomoon formation, prior to re-entering the Roche sphere (Panels \ref{fig:exomoon1}-\ref{fig:exomoon3}), and then the outcome of re-entry into the Roche sphere (Panels \ref{fig:exomoon4}-\ref{fig:exomoon6}). The \emph{pc\_earth} re-disruption is however very different. With a borderline Roche pericentre distance, it disrupts only a little at each subsequent passage. We observe 7 re-disruption cycles, in which its original mass is almost unchanged (see Section \ref{SS:Results_TidalStability} for further orbit analysis).

All other simulations in the upper part of Table \ref{tab:suite} (with the exception of \emph{velocity6-8}, see \ref{SS:Results_TidalStability}) generate a disc of debris, and the final exomoon mass is therefore subject to further considerations/modelling. Likewise, most of our new scenarios in the lower part of Table \ref{tab:suite}, generate disc masses that are nearly 0.2 M$_\mathrm{\oplus}$, or above. The \emph{bigger1-2} and \emph{biggest1-2} discs give rise to tidally stable exomoons (see \ref{SS:Results_TidalStability}), however the detectability of such moons with HEK remains to be analysed in Section \ref{SS:Results_Exomoons}, requiring a very large mass fraction of these discs to coagulate into an exomoon.

Analysis of the moonfall simulation data shows them to result in graze \& merge scenarios. Moonfalls lead to intact moons, retaining most of the mass of the original impactor, however due to dissipation in the impact, these moons return to either collide with the planet (\emph{moonfall1-2}) or tidally re-disrupt with a pericentre distance well within the planet's Roche limit (\emph{moonfall3-6}), in similarity to the \emph{velocity5} case.

\subsection{Tidal stability}\label{SS:Results_TidalStability}
We find that all impacts in the upper part of Table \ref{tab:suite}, apart from \emph{velocity6-8}, are stable to tidal interactions. The latter are hit-and-run scenarios, and the angular momentum carried away by the debris increases with the velocity, resulting in a slower planet rotation and therefore $r_\mathrm{sync}$ moves outwards, inhibiting stability. Had the planet been rotating, already prior to the impact, perhaps high-velocity collisions would have been able to produce tidally stable moons. Additionally, these hit-and-run scenarios naturally create sparse discs to begin with.

We also note that \emph{velocity5} and \emph{moonfall1-6} are all graze \& merge scenarios. Their exomoons emerge initially intact, however they do not survive subsequent collisions with the planet \emph{moonfall1-2} or tidal disruptions, as previously mentioned in Section \ref{SS:Results_DiscMass}. The only case that stands out as a graze \& capture scenario is \emph{pc\_earth}. Such a scenario creates an intact exomoon with relatively little debris besides it. Unlike all other simulations, whose exomoons coagulate from a disc, here the Roche radius does not imply where is the initial location of the satellite. Therefore stability has to be evaluated based on directly finding the orbit, which can be extracted from the simulation data. We cannot follow an extended post-impact orbit of this exomoons with SPH for computational-cost reasons, however for a simulation duration of 91 h, we managed to track the formation of this exomoon and additionally 7 more pericentre approaches (and minor disruptions). After each close approach we can calculate the pericentre distance based on the exomoon trajectory. If it remains to be larger than $r_\mathrm{roche}$, the exomoon is expected not to be stripped apart by tidal forces. If it is also larger than $r_\mathrm{sync}$, it implies long-term stability of subsequent tidal evolution. Our analysis of the data after the first pricentre approach shows the pericentre distance $q$, Roche limit $r_\mathrm{roche}$ and the synchronous orbit radius $r_\mathrm{sync}$ to be 15429 km, 14793 km and 9615 km, respectively. After the seventh pericentre approach $q=16602$ km, while $r_\mathrm{roche}$ and $r_\mathrm{sync}$ are almost unchanged. This exomoon is rapidly evolving outwards, and it had only lost about 2\% of its mass during these seven close approaches. We thus judge this exomoon to be tidally stable and survive its long-term evolution.

\subsection{Disc composition}\label{SS:Results_Composition}
Our results indicate that discs generated by the energetic impacts considered here for super-terrestrial planets, are often (a) iron-rich and (b) are composed mostly from impactor materials, although in about a third of the cases the impactor/target material fractions are almost identical. The former result is rather dissimilar to impacts around Earth-like planets, which noticeably generate extremely iron-poor discs, independent of whether SPH or AMR methods are used \citep{CanupEtAl-2013}. Impactor material fractions are however not so unlike.

In our entire suite of simulations the captured moon from \emph{pc\_earth} is interestingly also the richest in iron. With over 76\% iron in mass, it has a similar composition to planet Mercury, which also may have had much of its mantle stripped by a single impact or multiple impacts \citep{BenzEtAl-1988,ChauEtAl-2018}. It is also similar in terms of mass, with merely 50\% more than the mass of Mercury. We have identified a channel to form an Earth-exoplanet orbited by a Mercury-exomoon. 

\subsection{Exomoon properties}\label{SS:Results_Exomoons}
\begin{figure}
	\subfigure[SPH collision outcome (pre-switch)]{\label{fig:bigger2SPHend}\includegraphics[scale=0.4205]{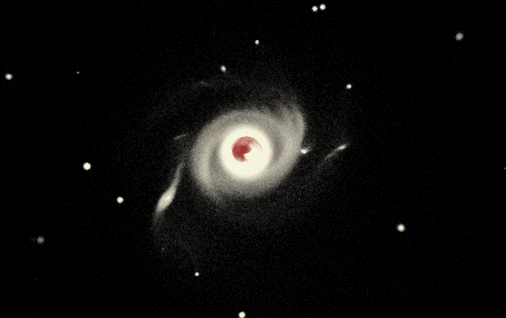}}
	\subfigure[N-body initial state]{\label{fig:bigger2NBodystart}\includegraphics[scale=0.4205]{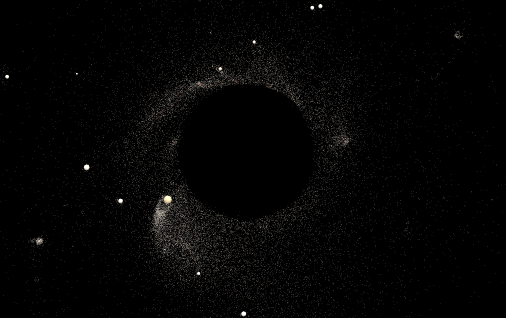}}
	\caption[r]{SPH-to-N-body handoff for the \emph{bigger2} simulation. (a) A top-down (semi-transparent) view of the post-collision SPH disc; fragments are identified, their properties (composition,origin) are calculated and handed over to (b) N-body simulation for long-term disc dynamics (inner Roche zone is free of particles); The colour scheme depicts composition, ranging from rocky material (white) to iron (red). Resolution: initial - $10^6$ SPH particles, follow-up - $\sim$30k disc fragments.}
	\label{fig:bigger2Handoff}
\end{figure}

In order to obtain the properties of the emerging exomoons we follow the long term dynamical gravitational interactions among the disc of debris, using an N-body code, as described in Section \ref{SS:NBody_Outline}. Figure \ref{fig:bigger2Handoff} shows an example of the hand over, for the \emph{bigger2} simulation. The colour scheme depicts composition, ranging from rocky material (white) to iron (red). The disc is predominantly composed of rock-dominated fragments (see also the composition in Table \ref{tab:suite}), which we identify immediately after the SPH is concluded. We also track the origin of the material. All this information is added to and processed in a modified REBOUND N-body code, providing us with detailed knowledge of the final assemblage of exomoons. As can be seen in Panel \ref{fig:bigger2NBodystart}, the inner Roche zone is devoid of particles during the N-body evolution, while mergers beyond the Roche zone lead to the coagulation of larger fragments over time. Since the inner Roche zone does not contribute mass to the disc in these simulations, what we eventually obtain is a lower limit mass.

\begin{figure}
	\subfigure[\emph{biggest2} scenario]{\label{fig:biggest2a_and_m}\includegraphics[scale=0.41]{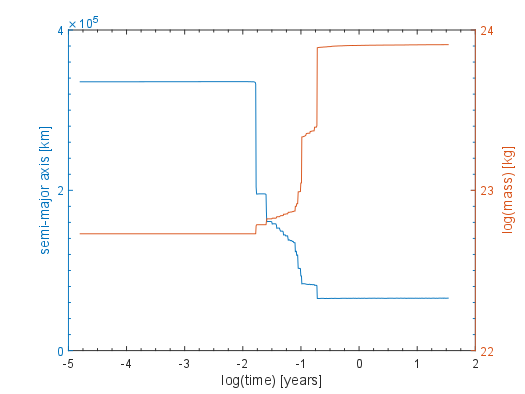}}
	\subfigure[\emph{gamma7} scenario]{\label{fig:gamma7a_and_m}\includegraphics[scale=0.41]{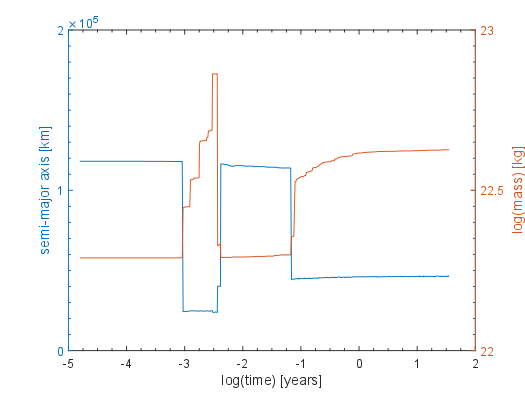}}
	\caption[r]{Examples of the temporal evolution of exomoon coagulation from an N-body simulation of a proto-satellite disc. We show the change in semi-major axis (left vertical axis) and mass (right vertical axis) of the largest fragment in the disc, for the \emph{biggest2} scenario (panel (a)) and the \emph{gamma7} scenario (panel (b)), respectively.}
	\label{fig:biggest2_and_gamma7}
\end{figure}

Analysis of the data shows that most of the mass at the end of the N-body simulations is concentrated in the two most-massive disc fragments. For example, in the \emph{bigger2} case, we start with $\sim30$k fragments and after 35 years of integration there remain only 188 fragments, wherein the two most-massive fragments constitute for 99\% of the mass in the disc. These numbers are typical to all of our simulations. Additionally, the mass ratio between the two most-massive fragments is always (except for one case, see Figure \ref{fig:gamma7a_and_m} and accompanying explanation) in the range 1:3-1:10.

Therefore, at the end of each simulation we compare the closest approach distance of the two-most massive fragments to their mutual Hill radius. We find that in no case does the former exceed the latter by more than a factor 2. Furthermore, configurations in which the closest approach is less than about four times greater than the mutual Hill radius are unstable \citep{ChatterjeeEtAl-2008}. Hence, we conclude that none of our configurations would ever result in two stable exomoons orbiting the planet. We then assume for simplicity, that we can merge the masses of the two most-massive fragments to obtain the mass of the exomoon, listed in column 12 of Table \ref{tab:suite}. We note that in principle, longer-term dynamics which include the mutual gravitational interactions of the remaining fragments, their tidal migration and the gravitational influence of the host star (in similarity with the \cite{CitronEtAl-2018} study), can also lead to ejection and de-orbiting, which have a higher probability to occur for the low mass fragment \citep{CitronEtAl-2018}. As previously mentioned in Section \ref{SS:NBody_Outline}, we avoid these additional complications which are not possible to account for without significant increase in the number of free parameters.

In Figures \ref{fig:biggest2_and_gamma7} we show two examples of the disc temporal evolution, by tracking the semi-major axis (left y-axis) and mass (right y-axis, log scale) of the largest fragment as a function of time (x-axis, log scale). In the \emph{biggest2} scenario (Panel \ref{fig:biggest2a_and_m}), discrete 'jumps' correspond to mergers, which increase the mass of the largest fragment and at the same time damp its semi-major axis. As can be seen, the largest fragment is initially 15 times less massive than at the end of the 35 year evolution. The eccentricity, not shown here, follows a similar trend to the semi-major axis. 

Unlike in the \emph{biggest2} scenario, where the largest fragment contains the overwhelming majority of the mass in the disc, and over an order of magnitude more mass than the 2nd largest fragment - the \emph{gamma7} scenario develops differently. The largest fragment starts to accrete and grow, as before, but approximately a day into the simulation its orbit, reduced by a previous merger, coincides with the planet's tidal sphere and consequently this fragment is omitted from the simulation. As the remaining fragments continue to grow, the end result is a disc which contains several large fragments, and the mass difference between the largest and 2nd largest fragments is not nearly as wide as it was in the \emph{biggest2} scenario or indeed any other scenario. The \emph{gamma7} scenario represents the exception rather than the rule.

Given our aforementioned calculation of exomoon masses after 35 years of evolution, the final exomoon masses in the upper part of Table \ref{tab:suite} show that 4 in 8 exomoons coagulate to form rocky satellites approximately twice more massive than any Solar system satellite. Together with the intact exomoon from \emph{pc\_earth}, we have 5 in 28 simulations (18\%) that generate final Mercury-like exomoons in the mass range of 0.04-0.08 $M_\oplus$. Interestingly they are generally less iron-rich than their respective proto-satellite discs, yet contain more material from the target. In the lower part of Table \ref{tab:suite}, all 4 simulated discs coagulate to generate exomoons that are approximately Mars-mass or above, however none exceed the HEK detection criteria. These exomoons generally have very similar material properties as their proto-satellite discs. At the end of our N-body simulations, all exomoons form at a semi-major axes of 1.2-2 times the Roche limit, similar to the initial relative formation distances for satellites around terrestrial planets \citep{ElserEtAl-2011}. Their eccentricities range between 0.05-0.3, with a rough correlation between $a$ and $e$ (more eccentric tend to be more outward). The only exception is the \emph{bigger1} case, in which $a=\sim 3 r_\mathrm{roche}$ and $e=0.5$.

\section{Discussion}\label{S:discussion}

\subsection{Comparison with previous studies}\label{SS:Comparison}
This work follows only one previous study which was recently carried out by BB17, and examined the possibility of forming exomoons detectable by HEK. Together, these are the only two studies ever performed which consider the formation of exomoons, given the highly energetic impacts relevant to super-terrestrial planets. In Section \ref{SS:Caveats} we argue that the details concerning the post-processing analysis, may lead to differences in the results between the two studies, which will not be related to the numerical method used, namely, SPH versus AMR respectively. We nevertheless observe that the disc masses follow precisely the same trend and that they vary by up to a factor of two, and typically much less, as can be directly compared in Table \ref{tab:suite}. The differences are more noticeable, and also more important for the most massive discs, because it is only the latter that have the potential to form exomoons which are potentially detectable using current technology. 

The most prominent example involves scenarios \emph{velocity5}, \emph{gamma3} and \emph{gamma4}. The latter have very similar collision parameters, and are expected to yield similar outcomes, as they do, in fact, in our study. In the study of BB17, however, the \emph{gamma3} disc is twice as massive as that of \emph{gamma4}. In Figure \ref{fig:gamma3Profile} which shows a snapshot of the density profile for the \emph{gamma3} post-impact planet, we discuss how the disc mass calculation is sensitive to how one models the planet's cross-section. Normally, we use a different analysis calculation compared to BB17 since we derive the appropriate scale-length directly from the data (due to the unorthodox shapes that sometimes emerge), rather than from a prescribed flattening (see Section \ref{SS:Caveats}). If we nevertheless change our algorithm to take the planet radius based on its material-derived density (in the same way we calculate it for the exomoons), our analysis outcome becomes a lot more similar to that of BB17. Hence, we see that the simulation method itself is sometimes not as important as the details of analysis, and even then differences between the two studies never amounted to more than a factor of two. A more accurate comparison requires (a) intimate knowledge of analysis details; (b) comparison between multiple parameters -- not only disc mass; and (c) visual inspection of data. It is therefore hard to manage if not performed by the same authors. Overall, we nevertheless conclude that the two methods broadly result in similar outcomes, somewhat validating the more accurate SPH versus AMR comparison study made by \cite{CanupEtAl-2013}.

The \emph{velocity5} disc mass differs the most between our study and that of BB17, however in this case we think the answer lies in the simulation duration. In our study we find that initially an intact moon is formed in the collision. It is more massive in our simulation than the disc mass given by BB17. We however predict based on its trajectory that it should return to tidally disrupt, and as previously mentioned (Section \ref{SS:Results_DiscMass}) we triple the fiducial simulation time to track this outcome. Indeed we find that the tidal disruption destroys the initially formed satellite, depositing most of the mass in the disc. This could readily explain the difference between the two studies.

Regarding tidal stability, BB17 do not provide analysis results for all their simulations. They mention only that \emph{pc\_earth} is marginally stable against planetary tides with $r_\mathrm{sync}\approx r_\mathrm{roche}$ and that \emph{velocity5} has $r_\mathrm{sync} \approx 2r_\mathrm{roche}$. In our study we identify these two cases to be the only graze \& merge/capture cases (besides the moonfall cases in the lower part of Table \ref{tab:suite}). Therefore, these are the only two cases wherein a comparison between $r_\mathrm{sync}$ and $r_\mathrm{roche}$ does not provide any information regarding tidal stability, because if there is no proto-satellite disc, the initial orbit of the satellite (if it formed) must be calculated directly, rather than from knowledge of $r_\mathrm{roche}$.

We also compare our conclusions regarding the question of detectability. Given the simplest HEK detection restriction, a currently-detectable exomoon requires a mass of at least 0.2 $M_\oplus$ (see the restrictive condition from \cite{KippingEtAl-2009}). BB17 set a more modest goal of identifying the collision phase space capable of generating debris discs more massive than 0.1 $M_\oplus$. By that restriction, they found that only 10 cases in a suite of 28 simulations were able to generate discs over 0.1 $M_\oplus$, and only 2 cases generated discs over 0.2 $M_\oplus$. Although they acknowledge that the disc mass can only serve as an upper limit on the final exomoon mass, they nevertheless conclude that "detectable rocky exomoons can be produced for a variety of impact conditions". As shown in Table \ref{tab:suite}, we have been able to both reproduce the BB17 results with similar outcomes, and also increase the number of discs with masses near or above 0.2 $M_\oplus$. Yet based on their results, as well as on ours, we actually arrive at exactly the opposite conclusion with respect to the HEK criteria. In the following Section (\ref{SS:Predictions}) we show that exomoons that coagulate from these disc would probably \emph{not} be detectable exomoons with present-day technology, but would nevertheless still be massive enough to be potentially detected with future instruments.

Finally, information about the origin of disc material is not provided in the study of BB17. The iron disc fraction is mention only for \emph{velocity5}. Without providing an exact fraction, they broadly suggest that the \emph{velocity5} disc is iron-rich. Our result is similarly an iron-rich disc, although, as previously mentioned, our interpretation of this disc is different.

\subsection{Detectable exomoons: predictions}\label{SS:Predictions}
The main goal of this paper was to predict the characteristics of exomoons that form through collisions among terrestrial and super-terrestrial planets. From the emerging trends in the data, we would also like to know which massive exomoons might be observable, either now (i.e., with HEK, given a mass over 0.2 $M_\oplus$) or in the future, given some improvement in our detection instruments.

In Section \ref{S:Formation} we show how studies of moon accretion restrict the fraction of mass in the disc that coagulates into a moon, to about 10-55\%. These studies were not performed for massive discs around super-Earths, and yet the chance to accrete an exomoon detectable by HEK from a disc which is only slightly more massive than 0.2 $M_\oplus$ appears marginal at best, even for the most optimistic assumptions. In this study we have hypothesized, based on the results of BB17, that the most likely avenues to generate massive discs are through Pluto-Charon type giant-impacts, scaled up to the super-Earth size range. We have tested this hypothesis and it turned out to indeed exclusively generate massive discs near or surpassing 0.2 $M_\oplus$. It is however not entirely clear how plausible or frequent such collisions are. They probably represent only a tiny fraction of the potential collision phase space during terrestrial planet formation. Moreover, since these discs never amount to more than 0.3 $M_\oplus$, the final mass of the exomoons generated by these discs should be in the range 0.03-0.15 $M_\oplus$ (based on previous studies which suggest $\sim$10-50\% of the disc mass coagulates to form moons). Hence, these projected masses are below the HEK detection criteria.

In order to further probe the final exomoon mass, we perform follow-up N-body simulations on the ensuing SPH discs, which now provide a lower-limit mass estimate. We study the coagulation of moons in discs initially larger than 0.1 $M_\oplus$, and find that half of the N-body simulations form exomoons that incorporate about a third of the disc mass. The other half however shows a remarkable diversity in the incorporation of disc material, the exomoon-to-disc mass ratio ranging between 0.2-98\%, with either very low or very high fractions. This result is interesting and very different to previous coagulation studies (see Section \ref{S:Formation}). We eventually obtain final exomoon masses that are diverse, in the range 0.002-0.186 $M_\oplus$. Hence, low limit mass estimates are still below, but close to the HEK detection criteria.

We also consider, for the first time, the possible formation of a massive exomoon through several mergers between smaller exomoons, instead of in a single giant-impact. Multiple impacts are a natural consequence of late terrestrial planet accretion, and have been considered previously as a possible formation mechanism for the Earth's moon \citep{CitronEtAl-2018}. Each impact forms a smaller moon which tidally evolves and gravitationally interacts with an existing, previously formed moon. With roughly equal probabilities, these moons can eject, merge, or fall back onto the planet. For the latter case, they typically fall at extremely grazing geometries with impact angle of the order of $\sim$90\textdegree~\citep{MalamudEtAl-2018}. Here we check whether such moonfalls lead to the destruction, ejection or survival of the impacting exomoons. We find that moonfalls result in graze \& merge scenarios, so while most of the exomoon mass survives the initial contact, its return trajectory takes it well within the planet's Roche limit, where it is expected to be almost entirely destroyed by tidal disruption. We conclude that moonfalls effectively restart the multiple-impact formation channel. Multiple-impact formation of massive moons is still possible, but it requires that the gravitational interaction between moonlets would exclusively result in mergers, while avoiding ejections and moonfalls, since both would terminate growth. Further studies are required in order to establish the relative probabilities of mergers, ejections and moonfalls, since the latter depend on the relative masses of the interacting moons, and since studies so far \citep{CitronEtAl-2018} have targeted the Earth rather than super-Earth planets. With more studies and statistics, we may be able to reach more decisive conclusions. 

According to this reasoning, we currently find no channel in which to form a sufficiently massive exomoon to be detectable by HEK. In the coming decade, only PLATO might enable a slight improvement, being able to detect Mars-like, $\sim$0.1 $M_{\oplus}$ exomoons (private correspondence with David Kipping, and also see \cite{RauerEtAl-2016}). In 4 of our N-body simulations we already confirm that such exomoons can form, however these formation channels entail large super-Earth collisions which probably represent only a tiny fraction of the potential collision phase space during terrestrial planet formation.

Unless we improve our ability to detect exomoons with future instruments, our results predict a difficulty in finding the first exomoon around any terrestrial or super-terrestrial planet. On the other hand, an additional factor $\sim$2 improvement (i.e. $\sim$0.05 $M_{\oplus}$, or about the mass of Mercury) in the detection threshold already makes exomoons far more likely to be detected. About a quarter of our simulations are compatible. We generate 8 exomoons close to or above this mass limit, in merely 12 follow-up N-body simulations of massive discs. We also form an additional Mercury-sized intact exomoon (\emph{pc\_earth}), directly after the initial collision, in a graze \& capture scenario. Our results indicate that these exomoons might be more iron-rich compared to solar system moons, and since they originate in a much broader collision phase space, they also represent more feasible outcomes of terrestrial planet formation.

Our study therefore motivates the development of a new generation of instruments, and sets a specific goal for its developers. Meanwhile, we suggest focusing our efforts with Kepler data on the biggest known Super-Earths, or else more massive planet categories. For the latter, future studies of exomoon formation are definitely required.


\subsection{Future studies}\label{SS:Future}
Our main suggestions for future studies are as follows:
\begin{itemize}
	\item Pre-rotation, collision geometries and resolution- both the BB17 study and our study consider initially non-rotating targets or impactors, as well as coplanar impact geometries, for simplicity. It is clear that initial rotation and an inclined collision plane would affect the disc masses and angular momenta. With improved high performance computing capabilities, future studies may choose to account for these complexities and consider a larger grid for the potential collision phase space. In Section \ref{SS:Outline}, we also motivate future resolution increase in similar studies, to corroborate our results.
	\item Dynamical and tidal evolution of multiple exomoons - as mentioned in Section \ref{SS:Predictions} (in the context of exomoon growth through multiple impacts) we require a detailed statistical understanding of gravitational interactions among exomoons, including their tidal evolution about their Super-terrestrial host planets. Such a study can easily follow the design of \cite{CitronEtAl-2018}, with the required modifications.
	\item Exomoon formation around gas-giant planets - our results suggest that exomoons may be more readily detected around more massive categories of planets. There is already one promising candidate, a Neptune-sized exomoon orbiting a super-Jupiter planet \citep{TeacheyKipping-2019}, although the data interpretation is controversial \citep{HellerEtAl-2019,KreidbergEtAl-2019,TeacheyEtAl-2019}. The work of \cite{HamersPortegiesZwart-2018} analytically calculates a tidal capture and subsequent orbital evolution scenario, however until now there are no studies performed to identify the detailed collision-formation of exomoons around this class of planets.	
\end{itemize}

\section{Acknowledgment}\label{S:Acknowledgment}
We thank the anonymous referee for helpful comments and suggestions that have improved the quality of our manuscript. UM and HBP acknowledge support from the Minerva center for life under extreme planetary conditions, the Israeli Science and Technology ministry Ilan Ramon grant and the ISF I-CORE grant 1829/12. CS and CB acknowledge support by the high performance and cloud computing group at the Zentrum f{\"u}r Datenverarbeitung of the University of T{\"u}bingen, the state of Baden-W{\"u}rttemberg through bwHP, the German Research Foundation (DFG) through grant no INST 37/935-1 FUGG, the FWF Austrian Science Fund project S11603-N16 and appreciate support by the DFG German Science Foundation project no. / 398488521.

	
\bibliographystyle{mnras} 
\bibliography{bibfile}     

\end{document}